\documentclass[a4paper,11pt]{article}
\pdfoutput=1 %

\usepackage{jcappub} %

\usepackage[T1]{fontenc} %

\title{Superfluid dark matter in tension with weak gravitational lensing data}

\author[a,b]{T. Mistele,\note{Corresponding author.}}
\author[b]{S. McGaugh,}
\author[a,c]{and S. Hossenfelder}

\affiliation[a]{Frankfurt Institute for Advanced Studies,\\Ruth-Moufang-Str. 1, 60438 Frankfurt am Main, Germany}
\affiliation[b]{Department of Astronomy, Case Western Reserve University,\\10900 Euclid Avenue, Cleveland, Ohio 44106, USA}
\affiliation[c]{Munich Center for Mathematical Philosophy, Ludwig-Maximilians-Universität,\\Geschwister-Scholl-Platz 1, 80539 München, Germany}

\emailAdd{tobias.mistele@case.edu}

\abstract{
    Superfluid dark matter (SFDM) is a model that promises to reproduce the successes of both particle dark matter on cosmological scales and those of Modified Newtonian Dynamics (MOND) on galactic scales.
    SFDM reproduces MOND only up to a certain distance from the galactic center, and only for kinematic observables: It does not affect trajectories of light.
    We test whether this is consistent with a recent analysis of weak gravitational lensing that has probed accelerations around galaxies to unprecedentedly large radii. This analysis found the data to be close to the prediction of MOND, suggesting they might be difficult to fit with SFDM.
    To investigate this matter, we solved the equations of motion of the model and compared the result to observational data.
    Our results show that the SFDM model is incompatible with the weak-lensing observations, at least in its current form.
}

\begin{document}
\maketitle
\flushbottom

\section{Introduction}
\label{sec:introduction}

Various models have recently been proposed that combine the successes of Modified Newtonian Dynamics (MOND) \cite{Milgrom1983a, Milgrom1983b, Milgrom1983c, Bekenstein1984} on galactic scales with those of $\Lambda$CDM on cosmological scales.
Examples are superfluid dark matter (SFDM) \cite{Berezhiani2015, Berezhiani2018}, the Aether Scalar Tensor (AeST) model \cite{Skordis2020, Skordis2021}, and the neutrino-based model $\nu$HDM \cite{Angus2009}.
In the following, we will focus on SFDM.

SFDM postulates a new type of particle that condenses to a superfluid around galaxies.
Inside this superfluid core of galaxies, the phonons of the superfluid mediate a force that, in a certain limit, resembles that of MOND.
At larger radii, the condensate breaks down and one matches the superfluid's density profile to that of a standard NFW halo at a matching radius $r_{\mathrm{NFW}}$ \citep{Berezhiani2018}.

Models like SFDM reproduce MOND only near the centers of galaxies, typically in the range where rotation curves are measured.
At larger radii, they deviate from MOND.
This is a desired feature for galaxy clusters where observations require accelerations larger than what MOND predicts \cite{Aguirre2001, Sanders2003}.
This deviation, however, can cause problems \cite{Mistele2023} when one also tries to fit recent weak-lensing data which has shown MOND-like behavior around galaxies up to $\sim1\,\mathrm{Mpc}$ \cite{Brouwer2021}.

SFDM in particular might conflict with this data for two reasons.
First, the MOND phenomena in SFDM are related to the condensed phase of a superfluid.
But at radii of about $\sim1\,\mathrm{Mpc}$, this condensed phase is expected to break down \cite{Berezhiani2015, Berezhiani2018, Hossenfelder2020}.
Second, even in the condensed phase, the MOND-like force of SFDM does not affect photons -- if it did, this would cause conflicts with the observations from GW170817 which saw the gravitational wave signal arrive at almost the same time as the electromagnetic signal \cite{Sanders2018, Boran2018, Hossenfelder2019}.
For this reason, gravitational lensing in SFDM works as in General Relativity (GR), just with the superfluid's mass as an additional source.
This means that even in the condensed phase, predictions for gravitational lensing in SFDM will differ from those of MOND.

More concretely, the acceleration inferred from kinematic observables includes both the acceleration due to the phonon force, $a_\theta$, and the standard Newtonian gravitational acceleration $a_N$.
Here, $a_N$ includes both the Newtonian acceleration due to the baryons, $a_b$, and that due to the superfluid's mass, $a_{\mathrm{SF}}$.
That is, $a_N = a_b + a_{\mathrm{SF}}$.
In contrast, the acceleration inferred from lensing observations does not include the phonon contribution $a_\theta$.
That is,
\begin{align}
\begin{split}
a_{\mathrm{obs}}\rvert_{\mathrm{kinematic}} &= a_N + a_\theta \,, \\
a_{\mathrm{obs}}\rvert_{\mathrm{lensing}} &= a_N \,.
\end{split}
\end{align}

Below we discuss whether or not SFDM can simultaneously explain kinematic and lensing observations, with a focus on the most recent weak lensing data \cite{Brouwer2021}.

\section{The superfluid dark matter model}
\label{sec:methods:model}

In the superfluid phase, the SFDM model is described by two fields, namely the Newtonian gravitational field $\phi_N$ and the phonon field $\theta$.
Instead of the Newtonian gravitational potential $\phi_N$, it will be convenient to use the quantities
\begin{align}
\begin{split}
\hat{\mu}(\vec{x}) &\equiv \mu_{\mathrm{nr}} - m \phi_N(\vec{x}) \,, \\
\varepsilon_*(\vec{x}) &\equiv \frac{2 m^2}{\alpha M_{\mathrm{Pl}} a_b(\vec{x})} \frac{\hat{\mu}(\vec{x})}{m} \,.
\end{split}
\end{align}
Here, $\mu_{\mathrm{nr}}$ is the superfluid's non-relativistic chemical potential, $m$ is the mass of the particles that make up the superfluid, $\alpha$ is a dimensionless model parameter, and $a_b$ is the Newtonian gravitational acceleration due to the baryons.
In spherical symmetry, $a_b = G M_b(r)/r^2$ with the enclosed baryonic mass $M_b$ and the Newtonian gravitational constant $G$.

We assume spherical symmetry and a point-particle baryonic density, $\rho_b = M_b \delta(\vec{x})$.
This is justified because the details of the baryonic mass distribution are not important at the large radii probed by weak lensing.

In spherical symmetry, the equation of motion for the phonon field $\theta$ reduces to an algebraic equation.
In particular, we have for the acceleration due to the phonon field $a_\theta$ \citep{Mistele2020, Mistele2022},
\begin{align}
\label{eq:atheta}
a_\theta = \sqrt{a_0 a_b} \cdot \sqrt{x_\beta(\varepsilon_*)} \,,
\end{align}
where $a_0$ and $\beta$ are further model parameters that we discuss below and $x_\beta$ is determined by a cubic equation,
\begin{align}
\label{eq:xeq}
0 = x_\beta^3 + 2 \left(\frac{2\beta}{3} - 1\right) \varepsilon_* \cdot x_\beta^2 + \left( \left(\frac{2\beta}{3}-1\right)^2 (\varepsilon_*)^2 - 1\right) x_\beta
- \left(\beta-1\right) \varepsilon_* \,.
\end{align}
Thus, $a_\theta$ can be calculated as an algebraic function of $a_b$ and $\varepsilon_*$.

The quantity $\varepsilon_*$ can be calculated from $\hat{\mu}$ and $a_b$, where $\hat{\mu}$ is determined by the usual Poisson equation, but with the superfluid's energy density $\rho_{\mathrm{SF}}$ as an additional source,
\begin{align}
\label{eq:muhateom}
\Delta\left(-\frac{\hat{\mu}}{m}\right) = 4 \pi G (\rho_b + \rho_{\mathrm{SF}}) \,.
\end{align}
The superfluid density can be written as \citep{Mistele2022}
\begin{align}
\label{eq:rhoSF}
\rho_{\mathrm{SF}} &= \frac23 \frac{m^2}{\alpha} \sqrt{a_0 a_b} M_{\mathrm{Pl}} \cdot
\frac{
  x_\beta(\varepsilon_*) (3-\beta) + 3(\beta-1) \varepsilon_*
}{\sqrt{
  x_\beta(\varepsilon_*) + (\beta-1) \varepsilon_*
}} \,.
\end{align}

Solving the Poisson equation requires a choice of boundary condition for $\hat{\mu}$.
Physically, this corresponds to a choice of chemical potential for the superfluid \citep{Mistele2019, Mistele2020}.
Thus, unlike in MOND, the solutions are not determined by the baryonic mass distribution alone.
They additionally depend on a choice of boundary condition ultimately set by galaxy formation.
Since galaxy formation in SFDM is not yet well-understood, we here treat this boundary condition as a free parameter for each galaxy.
See appendix~\ref{sec:appendix:numerical} for how we obtain solutions numerically.

The total acceleration felt by matter is that due to the Newtonian gravitational potential, $a_N$, plus that due to the phonon field, $a_\theta$,
\begin{align}
a_{\mathrm{tot}} = a_N + a_\theta = a_b + a_{\mathrm{SF}} + a_\theta \,,
\end{align}
where we split $a_N$ into a part due to the baryonic mass, $a_b$, and a part due to the superfluid, $a_{\mathrm{SF}}$.
In the so-called MOND limit
\begin{align}
\label{eq:mondlimit}
\varepsilon_* \ll 1 \,,
\end{align}
this total acceleration $a_{\mathrm{tot}}$ is close to the acceleration in MOND with an acceleration scale $a_0$ \citep{Berezhiani2015, Mistele2020},
\begin{align}
a_{\mathrm{tot}} = a_b + \sqrt{a_0 a_b} \,.
\end{align}
Specifically, this corresponds to MOND with an interpolation function $\nu(y) = 1 + 1/\sqrt{y}$.
In MOND, this interpolation function is not sufficient to match observations in general.
But here we are mostly interested in the large radii probed by weak lensing where all interpolation functions give $a_{\mathrm{tot}} \approx \sqrt{a_0 a_b}$.
Thus, for simplicity, we assume this interpolation function for MOND.

Unless otherwise stated, we use the fiducial numerical parameter values proposed by the original authors of the model \cite{Berezhiani2018}, namely $m^2/\alpha = 0.18\,\mathrm{eV}^2$, $\beta=2$, and $a_0 = 0.87\cdot 10^{-10}\,\mathrm{m}/\mathrm{s}^2$.
This fiducial value of $a_0$ is a bit lower than the best-fit value in MOND, namely $a_0^{\mathrm{MOND}} \approx 1.2 \cdot 10^{-10}\,\mathrm{m}/\mathrm{s}^2$ \citep{Lelli2017b}.
The reason for choosing this smaller value is to compensate for possible contributions from $a_{\mathrm{SF}}$.
We argue in appendix~\ref{sec:appendix:parameters} that changing the numerical parameter values does not qualitatively affect our conclusions.

For concreteness, when comparing to MOND, we always use the larger value $a_0^{\mathrm{MOND}}$ for MOND, not the SFDM value $a_0$,
\begin{align}
\label{eq:aMOND}
a_{\mathrm{MOND}} \equiv a_b + \sqrt{a_0^{\mathrm{MOND}} a_b} \,.
\end{align}

As already mentioned above, the total acceleration $a_{\mathrm{tot}} = a_\theta + a_N$ is relevant only for kinematic observables like galactic rotation curves.
The phonon field does not contribute to gravitational lensing because light does not directly couple to the new field.
As a result, lensing works as in $\Lambda$CDM just with the superfluid mass instead of the cold dark matter mass \citep{Hossenfelder2019}.
That is, if one uses the same formalism as in $\Lambda$CDM to infer an acceleration around galaxies from lensing, this inferred acceleration corresponds to $a_N$ in SFDM.

\section{Deviations from MOND for kinematic observables}
\label{sec:kinematic}

We first consider kinematic observables, like rotation curves, on their own.
Since kinematic observations are well-explained by MOND, we first quantify the deviation of SFDM from MOND and identify the solutions with minimal deviations.
We then compare these optimal, and close-to-optimal, solutions to observations.

By solutions we here mean solutions of the SFDM equations of motion.
Optimal and non-optimal solutions differ from each other in the boundary condition that one must choose when solving these equations.
Physically, this means the superfluids of different solutions have different chemical potentials, see section~\ref{sec:methods:model}.

The total acceleration felt by matter in SFDM follows the MOND prediction only at sufficiently small galactocentric radii.
At larger radii, it deviates from MOND.
Where exactly these deviations set in and how large they are depends on the model parameters, the boundary condition, and the baryonic mass of the galaxy.

Specifically, consider a galaxy with a fixed baryonic mass $M_b$ and suppose we allow the acceleration in SFDM to deviate from that in MOND by at most a fraction $\delta$.
Then, there is an optimal boundary condition for which this condition is fulfilled up to the maximum possible radius $r_{\mathrm{max}}$.

\begin{figure}
\centering
\includegraphics[width=.8\textwidth]{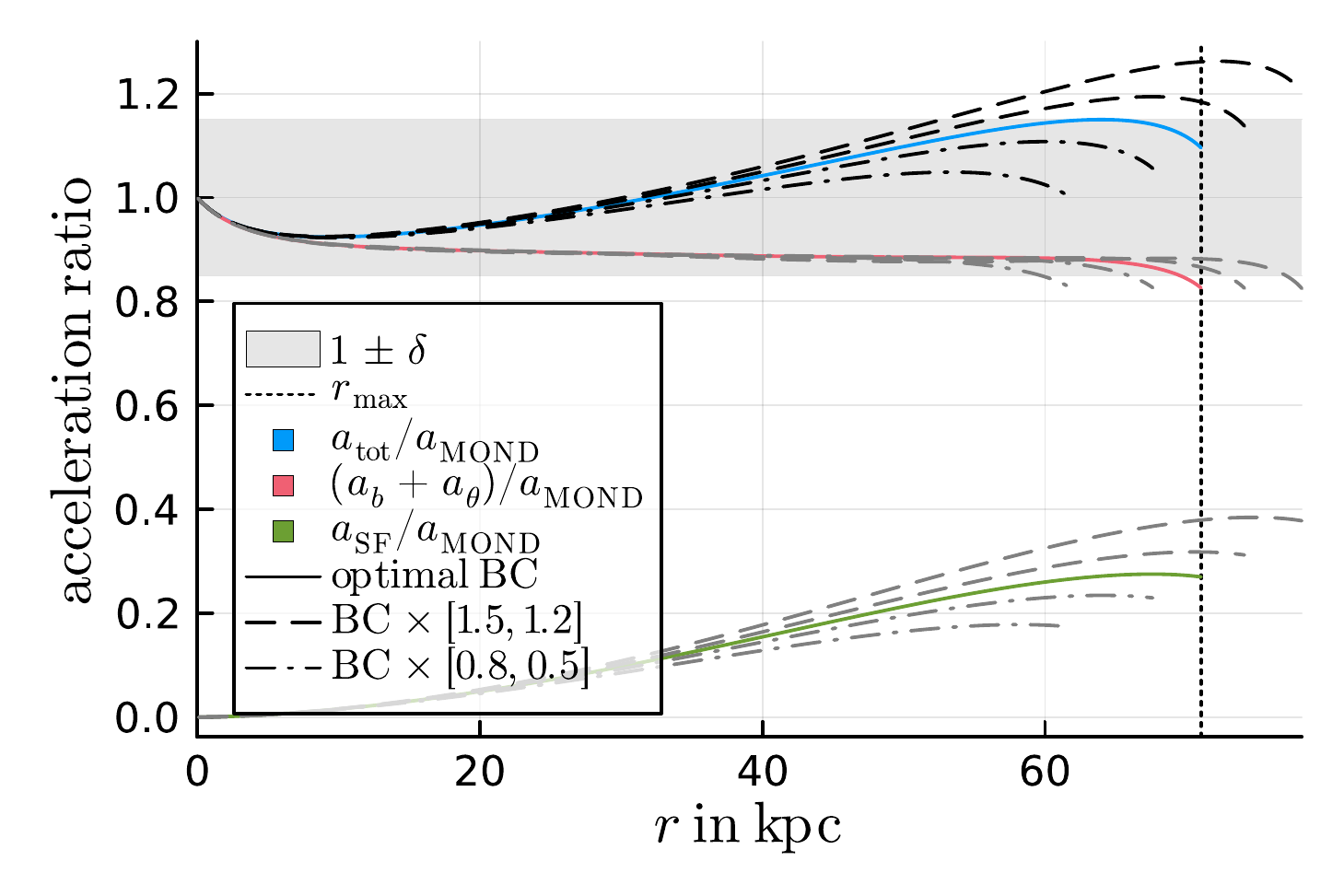}
\caption{
  The total kinematically-inferred acceleration $a_{\mathrm{tot}} = a_b + a_{\mathrm{SF}} + a_\theta$ inside the superfluid core (solid blue line) relative to the acceleration in MOND for an example galaxy with $M_b = 2\cdot 10^{10}\,M_\odot$.
  This is for the boundary condition for which $a_{\mathrm{tot}}$ stays within a fraction $\delta = 0.15$ from MOND up to the maximum possible radius $r_{\mathrm{max}}$ (vertical dotted line).
  Dashed (dash-dotted) black lines show solutions with this optimal boundary condition multiplied by $1.5$ and $1.2$ ($0.8$ and $0.5$).
  Also shown are the different components that make up $a_{\mathrm{tot}}$, namely the baryonic Newtonian acceleration plus the acceleration due to the phonon field, $a_b + a_\theta$, (solid red line) and the superfluid's Newtonian acceleration $a_{\mathrm{SF}}$ (solid green line).
  For the non-optimal boundary conditions, $a_b + a_\theta$ and $a_{\mathrm{SF}}$ are shown as grey lines.
  All solutions are shown up to the radius where the corresponding superfluid density $\rho_{\mathrm{SF}}$ drops to zero.
}
\label{fig:illustrate-deltatot}
\end{figure}

We illustrate this in Fig.~\ref{fig:illustrate-deltatot} for one example galaxy with $M_b = 2 \cdot 10^{10}\,M_\odot$.
For the optimal boundary condition (solid blue line), the total kinematically-inferred acceleration in SFDM stays within $15\%$ from MOND up to a maximum radius $r_{\mathrm{max}}$ of about $70\,\mathrm{kpc}$.
For all other boundary conditions (dashed and dash-dotted black lines), either deviations from MOND become larger than $15\%$ before $70\,\mathrm{kpc}$ or the condensate phase ends before $70\,\mathrm{kpc}$.
While this example illustrates a typical case, the precise numerical values will be different for different galaxies.

That the condensate phase ends means that the condensate density drops to zero, $\rho_{\mathrm{SF}} = 0$.
Mathematically, the solutions can be continued beyond this point but the physical meaning of this regime is unclear. Most likely, the negative density indicates an instability, so we truncate all solutions at $\rho_{\mathrm{SF}} = 0$.

\begin{figure}
\centering
\includegraphics[width=.8\textwidth]{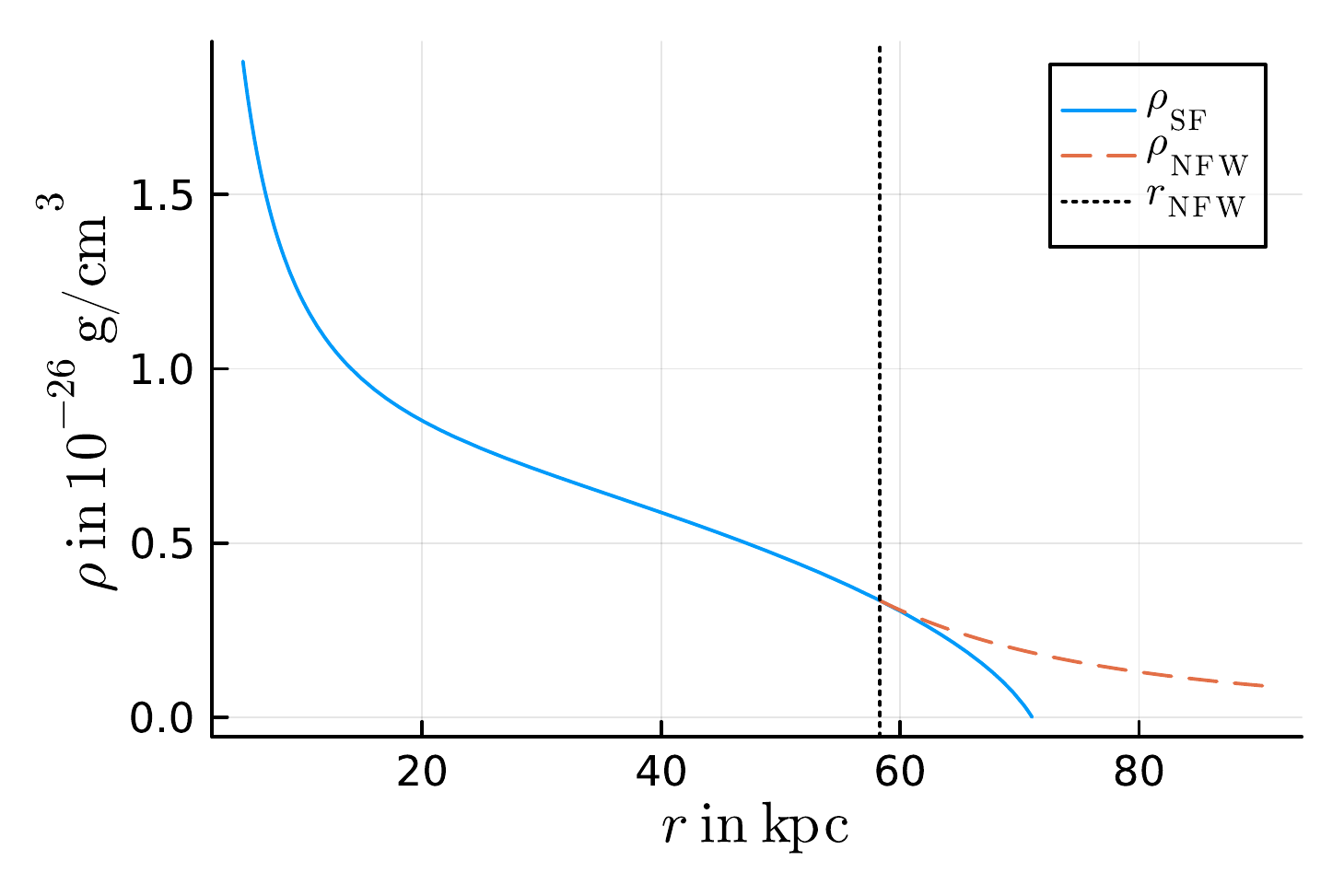}
\caption{
  The superfluid density $\rho_{\mathrm{SF}}$ (solid blue line) for the solution with $\delta = 0.15$ for an example galaxy with $M_b = 2 \cdot 10^{10}\,M_\odot$.
  The dashed red line illustrates one possibility of matching an NFW tail, $\rho_{\mathrm{NFW}} \propto 1/r^3$, to the superfluid density $\rho_{\mathrm{SF}}$.
  Here, we show the matching radius $r_{\mathrm{NFW}}$ that gives the largest dark matter acceleration $a_{\mathrm{DM}}$ outside the superfluid core, but other choices are also possible (see for example Fig.~\ref{fig:SFDM-illustrate-kinematic-vs-lensing-NFWtail}).
}
\label{fig:illustrate-rhoSF-rNFW}
\end{figure}

In SFDM, it is usually assumed \cite{Berezhiani2015, Berezhiani2018} that the condensate phase transitions into a non-condensed cold dark matter phase already before $\rho_{\mathrm{SF}} = 0$.
In particular, the idea is that the superfluid density is matched to an NFW halo at a radius $r_{\mathrm{NFW}}$.
This is illustrated for one example galaxy in Fig.~\ref{fig:illustrate-rhoSF-rNFW}.

Here, we treat the matching radius $r_{\mathrm{NFW}}$ as a free parameter for each galaxy since the transition from the condensed to the not-condensed phase is still poorly understood \citep{Mistele2020}.
To keep things simple, we use a $1/r^3$ NFW tail instead of a full NFW profile.
We do not expect this to qualitatively change the results \citep{Hossenfelder2019}.

In principle, how long the kinematically-inferred acceleration stays close to MOND could depend on where and how the superfluid core is matched to the NFW tail.
However, we consider only the condensate without any NFW matching for our definition of the maximum radius $r_{\mathrm{max}}$ and the corresponding optimal boundary condition.
That is, Fig.~\ref{fig:illustrate-deltatot} does illustrate the actual definitions we use although it does not involve any matching to an NFW halo.
There are two reasons for using this definition.

First, the condensate core is what matters most for comparing SFDM to MOND.
This is because the relation of SFDM to MOND comes from the phonon force which acts only inside the condensate.
Outside, the total acceleration will in general be different from MOND.

Of course, in principle, SFDM can match the MOND prediction even outside the condensate.
But this is possible only by carefully adjusting a boundary condition separately for each galaxy.
This is similar to the situation in $\Lambda$CDM.
One of the main motivations behind SFDM is precisely to reproduce MOND without such tuning.
Thus, SFDM loses one of its main advantages if we allow to match MOND by tuning the boundary condition.

Second, the maximum radius and the optimal boundary condition often don't change when taking into account a transition to an NFW halo.
The reason is that the phonon force does not act outside the superfluid core so that, at the transition radius $r_{\mathrm{NFW}}$, the kinematically-inferred acceleration drops from $a_\theta + a_N$ to just $a_N$.
Often, $a_\theta$ is the most important contribution, see for example Fig.~\ref{fig:illustrate-deltatot}.
When it drops out, the acceleration typically immediately deviates from MOND by more than a fraction $\delta$ at the transition radius $r = r_{\mathrm{NFW}}$.
The optimal transition radius is then just where $\rho_{\mathrm{SF}}$ drops to zero.
Thus, the optimal boundary condition is the same with and without taking into account a transition to an NFW tail.

\begin{figure}
\centering
\includegraphics[width=.8\textwidth]{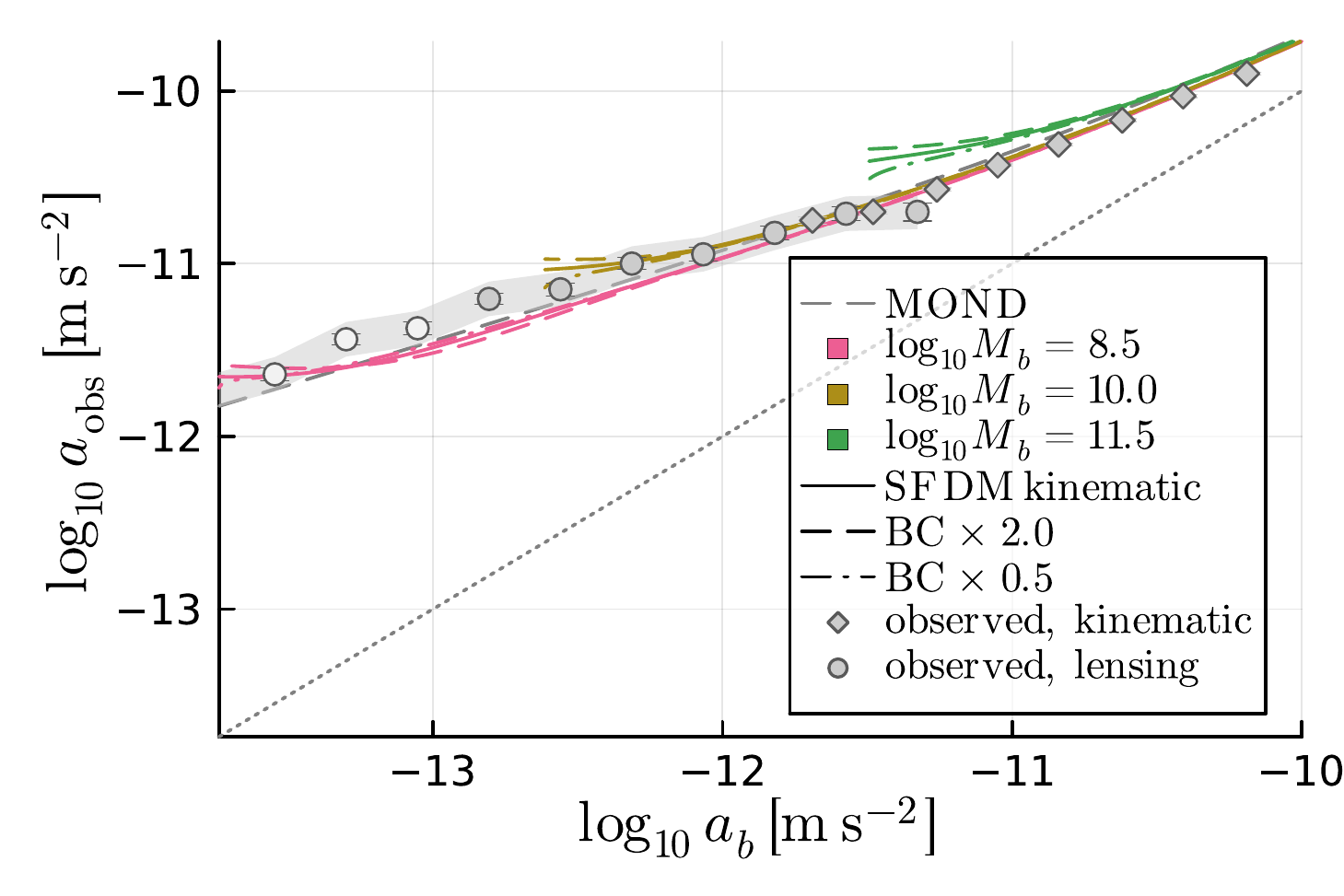}
\caption{
  The total acceleration $a_{\mathrm{obs}}$ that would be inferred from kinematic observations for numerical solutions in SFDM as a function of $a_b$ for various baryonic masses and boundary conditions.
  Solid lines show, for each $M_b$, the optimal boundary condition for $\delta = 0.8$.
  That is, the boundary condition for which the kinematically-inferred $a_{\mathrm{obs}}$ stays closer than a fraction $\delta$ to MOND up to the maximum possible radius.
  Solutions with this boundary condition multiplied by $0.5$ and $2.0$ are shown as dash-dotted and dashed lines, respectively.
  These solutions are valid inside the superfluid cores of galaxies.
  We do not continue them with an NFW tail (see Fig.~\ref{fig:SFDM-illustrate-kinematic-vs-lensing-NFWtail} for that).
  For a given $M_b$, solutions are cut off where the first of the solutions reaches zero superfluid density.
  Also shown are the observed kinematic RAR \cite{Lelli2017b} and the observed lensing RAR \cite{Brouwer2021}.
  The lensing data does not include the hot gas estimate used in some places in the original weak lensing analysis \cite{Brouwer2021}.
  We correct the observed weak-lensing RAR to be consistent with the $M/L_*$ scale of the observed kinematic RAR (McGaugh 2022, priv. comm.).
  Data points below $a_b = 10^{-13}\,\mathrm{m/s}^2$ have a lighter color since the isolation criterion assumed in  the weak lensing analysis \cite{Brouwer2021} is less reliable there.
  The error bars indicate the statistical uncertainties and the gray band indicates the $0.1\,\mathrm{dex}$ systematic uncertainty from the excess surface density to RAR conversion \cite{Brouwer2021}.
}
\label{fig:SFDM-illustrate-kinematic}
\end{figure}

In MOND, there is a one-to-one relation between the observed acceleration $a_{\mathrm{obs}}$ and the baryonic Newtonian acceleration $a_b$.
In particular, MOND predicts $a_{\mathrm{obs}} = a_b$ at large accelerations, $a_b \gg a_0$, and $a_{\mathrm{obs}} = \sqrt{a_0 a_b}$ at small accelerations, $a_b \ll a_0$ (Eq.~\eqref{eq:aMOND}).
Kinematic observations confirm this relation, see for example the Radial Acceleration Relation (RAR) \cite{Lelli2017b}.

In SFDM, there is no longer such a one-to-one relation.
This is illustrated in Fig.~\ref{fig:SFDM-illustrate-kinematic} which shows the relation between $a_{\mathrm{obs}}$ and $a_b$ for galaxies with various baryonic masses and boundary conditions.
We see that there is a strong trend with baryonic mass:
Smaller galaxies more easily follow the MOND prediction than larger galaxies.
This was first noticed in Ref.~\citep[appendix A.1.2]{Mistele2022} and is theoretically explained in appendix~\ref{sec:appendix:estar}.
This trend may make SFDM less attractive compared to MOND, but is by itself not enough to rule out SFDM using kinematic data.
For a quantitative discussion of fitting kinematic data with SFDM, we refer the reader to Ref.~\citep{Mistele2022}.

\section{Deviations from MOND for weak lensing}
\label{sec:weaklensing}

\begin{figure}
\centering
\includegraphics[width=.8\textwidth]{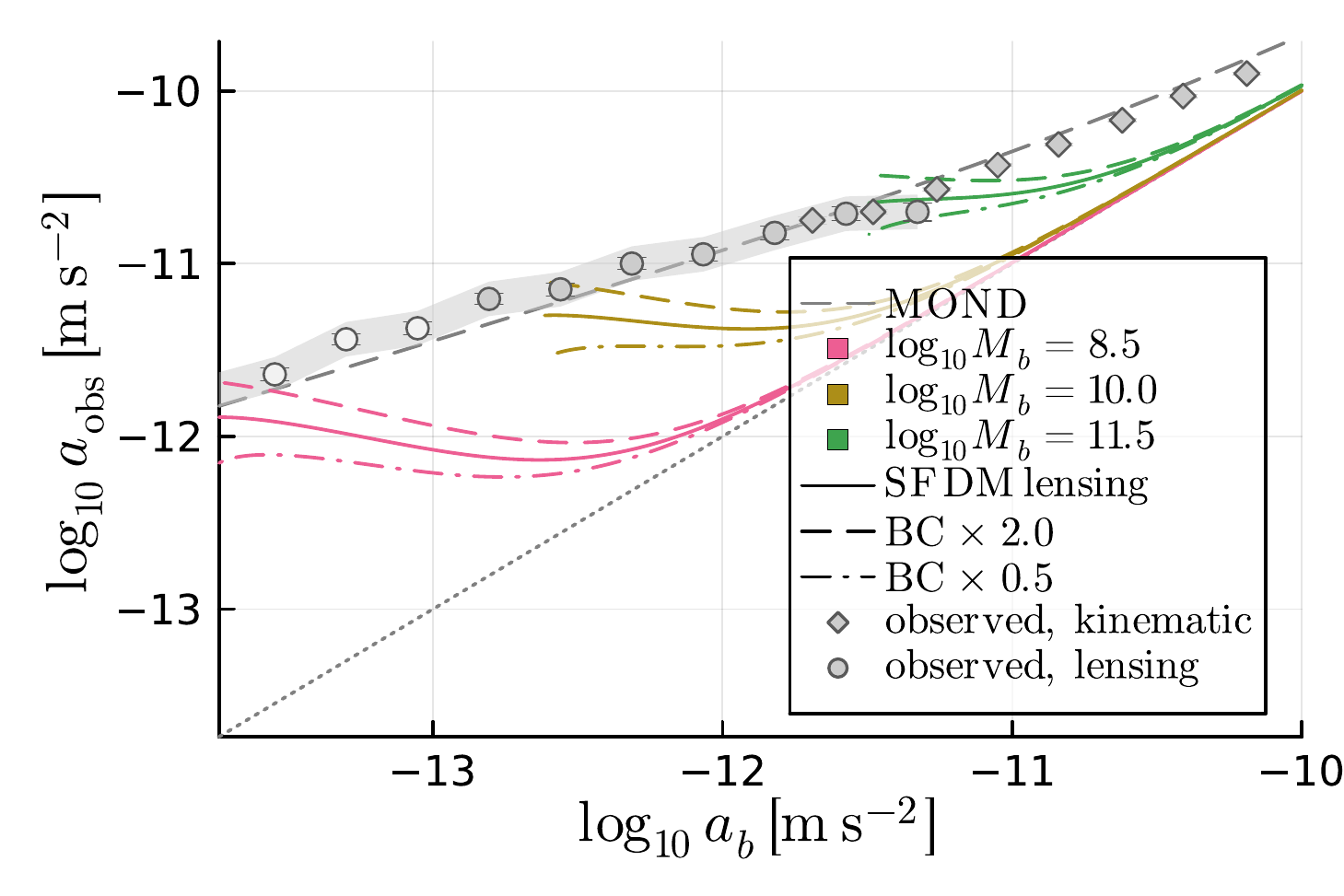}
\caption{
  As Fig.~\ref{fig:SFDM-illustrate-kinematic} but for the lensing-inferred $a_{\mathrm{obs}}$ instead of the kinematically-inferred $a_{\mathrm{obs}}$.
  These differ in SFDM because the phonon acceleration $a_\theta$ contributes only to kinematic observables.
}
\label{fig:SFDM-illustrate-lensing}
\end{figure}

As already mentioned above, photons are not coupled to the superfluid's phonons in order to avoid constraints from GW170817.
Thus, lensing works as in $\Lambda$CDM just with the superfluid's mass as an additional source.
In particular, the acceleration inferred from lensing is not $a_N + a_\theta$ but just $a_N$.

The consequences for the RAR are illustrated in Fig.~\ref{fig:SFDM-illustrate-lensing}.
In contrast to the kinematic RAR shown in Fig.~\ref{fig:SFDM-illustrate-kinematic}, the lensing RAR almost never follows the MOND prediction.
Even when carefully adjusting the boundary condition, the lensing RAR of SFDM follows the MOND prediction only in a very small spatial region, not over an extended range of radii.
This is in tension with lensing observations \cite{Brouwer2021} that agree well with the MOND prediction.

\begin{figure}
\centering
\includegraphics[width=.8\textwidth]{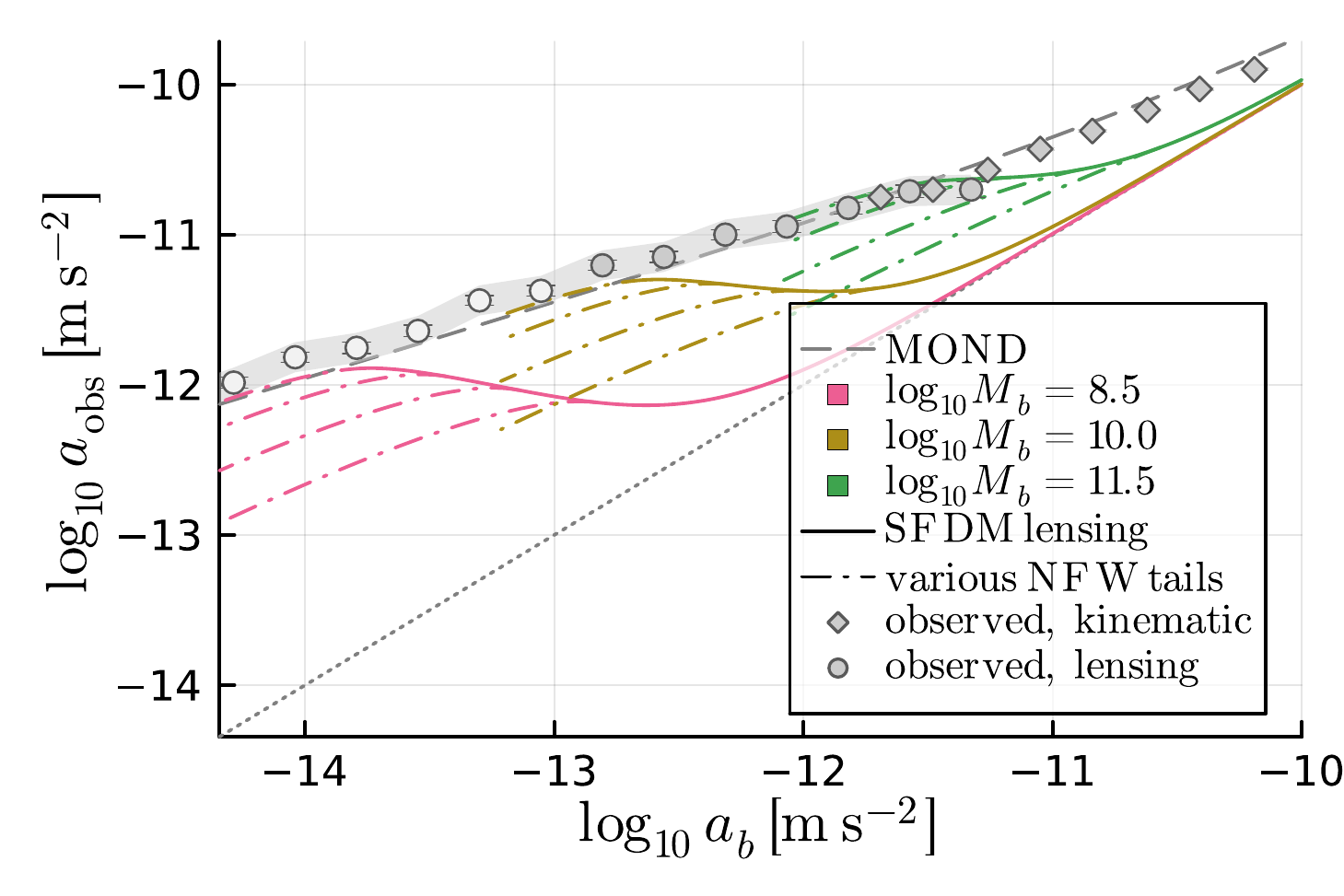}
\caption{
  As Fig.~\ref{fig:SFDM-illustrate-lensing}, but for a fixed boundary condition corresponding to $\delta = 0.8$ and with the superfluid density matched to an $1/r^3$ NFW tail at various radii $r_{\mathrm{NFW}}$.
  Solid lines show the lensing-inferred $a_{\mathrm{obs}}$ inside the superfluid core, at $r < r_{\mathrm{NFW}}$.
  Dash-dotted lines show $a_{\mathrm{obs}}$ outside the superfluid, at $r > r_{\mathrm{NFW}}$.
  For each $M_b$, we show $a_{\mathrm{obs}}$ for the $r_{\mathrm{NFW}}$ that gives the maximum possible lensing-inferred $a_{\mathrm{obs}}$ outside the superfluid.
  We also show NFW tails with $0.35$, $0.5$, and $0.7$ times this optimal $r_{\mathrm{NFW}}$.
  Outside the superfluid, at $r > r_{\mathrm{NFW}}$, the kinematic and the lensing $a_{\mathrm{obs}}$ coincide since the phonon force does not act there \citep{Berezhiani2015}.
}
\label{fig:SFDM-illustrate-kinematic-vs-lensing-NFWtail}
\end{figure}

Both Fig.~\ref{fig:SFDM-illustrate-lensing} and Fig.~\ref{fig:SFDM-illustrate-kinematic} show only the superfluid core around a galaxy.
But as discussed above, this superfluid core ends at a finite radius $r_{\mathrm{NFW}}$.
At this radius, the superfluid density is matched to that of a standard NFW halo and the phonon force is no longer active, not even for kinematic observables.
We show the lensing RAR including such an NFW tail for various choices of $r_{\mathrm{NFW}}$ in Fig.~\ref{fig:SFDM-illustrate-kinematic-vs-lensing-NFWtail}.
We see that introducing an NFW tail allows the SFDM lensing RAR to more easily follow the MOND prediction.
That is, outside the superfluid core, the shape of the lensing-inferred $a_{\mathrm{obs}}$ is much closer to the MOND prediction.
Of course, asymptotically, the MOND acceleration scales as $1/r$ while that of the NFW tail scales as $\ln(r)/r^2$ which is very different.
Still, as Fig.~\ref{fig:SFDM-illustrate-kinematic-vs-lensing-NFWtail} demonstrates, an NFW tail can reasonably reproduce the shape of the MOND acceleration over a non-negligible range of radii.

However, a big problem remains: For the normalization -- not just the shape -- to come out correctly, one has to carefully choose the boundary condition and matching radius $r_{\mathrm{NFW}}$ separately for each galaxy.
For this SFDM has to rely on galaxy formation to pick the correct values.
But the goal of SFDM was precisely to be able to reproduce the MOND prediction without having to rely on galaxy formation.
So this is not satisfactory.

So far, we have considered the lensing or kinematic signal of a single galaxy in SFDM.
The observed weak-lensing RAR \cite{Brouwer2021}, however, is available only in an averaged sense for a large number of stacked galaxies.
Thus, accelerations larger than MOND from one galaxy may cancel with accelerations smaller than MOND from another galaxy, giving a stacked lensing RAR that is more MOND-like than the lensing RAR of any individual galaxy.
This gives a bit more freedom in the boundary conditions and NFW matching radii that galaxy formation needs to pick for SFDM to reproduce the observed weak-lensing RAR.
However, it's still far from clear why galaxy formation would pick just the right boundary conditions and matching radii to produce the desired cancellations.

Another problem for SFDM is that observations continue without gap from the kinematic to the lensing RAR (see the data points in, for example, Fig.~\ref{fig:SFDM-illustrate-kinematic}).
In SFDM, as illustrated by Fig.~\ref{fig:SFDM-illustrate-kinematic}, Fig.~\ref{fig:SFDM-illustrate-lensing}, and Fig.~\ref{fig:SFDM-illustrate-kinematic-vs-lensing-NFWtail}, the lensing RAR can be close to the MOND prediction only when the kinematic RAR is not and vice versa.

In principle, with special boundary conditions and special galaxy samples, it could be that lensing observations never see the relatively small galactocentric radii of galaxies where the lensing-inferred $a_{\mathrm{obs}}$ is much smaller than in MOND while kinematic observations never reach sufficiently large radii where the kinematically-inferred $a_{\mathrm{obs}}$ is significantly larger than in MOND.
In this case, SFDM might match both kinematic and lensing observations.
But this seems highly unlikely.

In some cases,  we can make an even stronger statement about the tension between the lensing and kinematic RAR in SFDM.
To see this, suppose we require the kinematic RAR to follow the MOND prediction within the superfluid core (where the phonon force is active).
This gives an upper bound on the lensing-inferred $a_{\mathrm{obs}}$ outside the superfluid core (see appendix~\ref{sec:appendix:aDMupperbound}).
In some cases, this upper bound is smaller than what lensing observations imply.
In these cases, no amount of tuning the boundary condition and NFW matching radius can make the lensing RAR in SFDM agree with observations.

As an example, consider a fixed baryonic mass $M_b$.
Consider then the optimal solution that, kinematically, stays within a fraction $\delta = 0.8$  of the MOND prediction up to the maximum possible radius $r_{\mathrm{max}}$.
Given this kinematically optimal solution, there is an optimal matching radius $r_{\mathrm{NFW}}$ which produces the largest value for the lensing-inferred $a_{\mathrm{obs}}$, see appendix~\ref{sec:appendix:aDMupperbound}.
This optimal matching radius $r_{\mathrm{NFW}}$ and the corresponding maximal lensing-inferred $a_{\mathrm{obs}}$ are illustrated in Fig~\ref{fig:SFDM-illustrate-kinematic-vs-lensing-NFWtail}.
We see that even for this optimal matching radius $r_{\mathrm{NFW}}$, the maximal lensing-inferred $a_{\mathrm{obs}}$ only barely reaches the MOND prediction.
Repeating the same procedure with a kinematically-optimal solution for a smaller allowed deviation from MOND, i.e. a smaller $\delta$, the lensing-inferred $a_{\mathrm{obs}}$ often stays significantly below the MOND prediction, no matter which matching radius $r_{\mathrm{NFW}}$ is chosen.

\subsection{Quantitative analysis}
\label{sec:weaklensing:quantitative}

Finally, we illustrate some of the qualitative points we made above by a simple quantitative analysis.
For each galaxy, we minimize the reduced $\chi^2$ as a function of two parameters, namely the boundary condition and the NFW matching radius.
We have
\begin{align}
 \chi^2_{\mathrm{red}} \equiv \frac{1}{N - N_{\mathrm{dof}}} \sum_i \frac{ ( a_{\mathrm{obs,SFDM},i} - a_{\mathrm{obs,lensing},i})^2 }{\sigma_i^2} \,,
\end{align}
where $N = 15$ is the number of weak-lensing data points, $N_{\mathrm{dof}} = 2$ is the number of fit parameters,  $i$ runs over the $N$ data points, $a_{\mathrm{obs,lensing},i}$ is the $i$-th weak-lensing-inferred acceleration with statistical uncertainty $\sigma_i$, and $a_{\mathrm{obs,SFDM},i}$ is the $i$-th $a_{\mathrm{obs}}$ value calculated in SFDM.
We minimize this reduced $\chi^2$ using the Julia package `Optim.jl` \citep{Mogensen2018}.

\begin{figure}
\centering
\includegraphics[width=.8\textwidth]{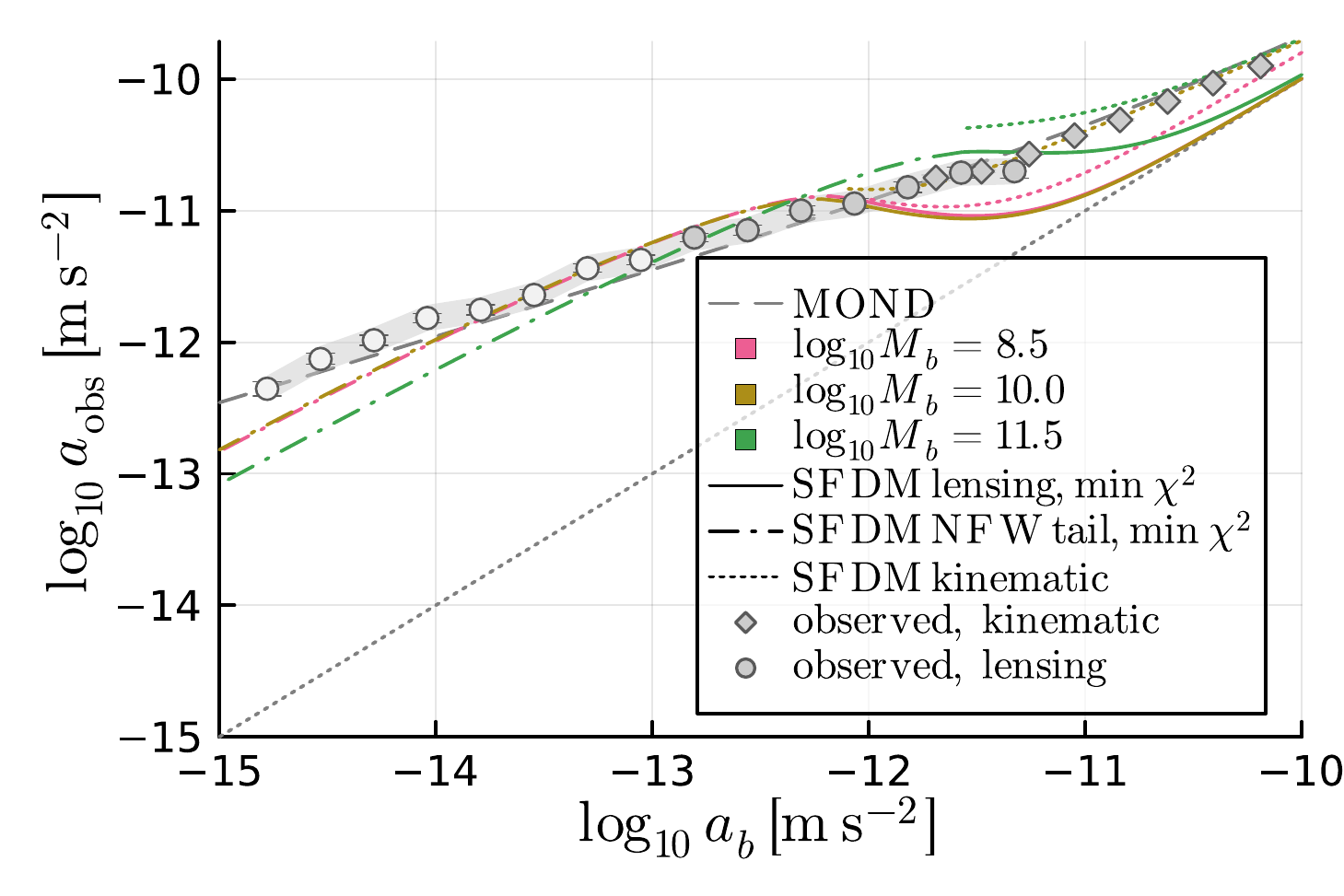}
\caption{
  As Fig.~\ref{fig:SFDM-illustrate-kinematic-vs-lensing-NFWtail} but for the ``closest-match'' boundary condition and NFW matching radius for each $M_b$.
  In addition, we show the kinematically-inferred acceleration for each $M_b$ (dotted lines).
  We use the term ``closest match'' rather than ``best fit'' for the solutions with minimum reduced $\chi^2$ to indicate that none of these solutions match all of the data.
  This is explained in more detail in the main text.
  As also explained in the main text, the closest-match solutions for $\log_{10} M_b/M_\odot = 8.5$ and $10.0$ represent the ``universal'' best way to fit the lensing data in SFDM.
  Galaxies with larger masses, such as $\log_{10} M_b/M_\odot = 11.5$, cannot reach this ``universal'' closest match.
}
\label{fig:SFDM-illustrate-best-fit}
\end{figure}

In Fig.~\ref{fig:SFDM-illustrate-best-fit} we show solutions for the ``closest-match'' parameters for three galaxies with $\log_{10} M_b/M_\odot$ values of $8.5$, $10.0$, and $11.5$.
We call these ``closest-match'' rather than ``best-fit'' because none of the solutions match all of the data; these are the least bad results we can obtain.
The corresponding minimized $\chi_{\mathrm{red}}^2$ values are $15.0$, $14.9$, and $28.7$.
These are worse than what we find in MOND with no free parameters, $N_{\mathrm{dof}} = 0$, namely $\chi^2_{\mathrm{red}} = 6.5$.
However, one should not read too much into the precise numerical values here.
Formally, all of these $\chi^2_{\mathrm{red}}$ values indicate bad fits, since they are all much larger than $1$.
But that does not mean that the corresponding models are all ruled out.
Indeed, the systematic uncertainties -- such as those in the stellar masses and in converting the excess surface density profiles to accelerations -- are larger than the statistical uncertainties that enter the $\chi^2$ calculation \citep{Brouwer2021}.
The situation is similar for kinematic observations, see for example \citep{Li2021}.

It is more informative to look at the qualitative features of the closest-match solutions shown in Fig.~\ref{fig:SFDM-illustrate-best-fit}.
For example, we see that the closest-match solutions for $\log_{10} M_b/M_\odot = 8.5$ and $10.0$ are very similar.
Indeed, all closest-match solutions would ideally follow exactly this ``universal'' closest match to the lensing data.
That is, all SFDM closest-match solutions would ideally match to an NFW tail at about $a_b \sim 10^{-12}\,\mathrm{m/s}^2$ -- slightly after the first lensing data point -- with an $a_{\mathrm{obs}}$ slightly larger than the data points.
This is the ``universal'' best way to fit the lensing data in SFDM because the superfluid itself does not match the lensing data at all (Fig.~\ref{fig:SFDM-illustrate-lensing}) and an NFW tail does not have the right scaling (namely $1/r^3$ compared to the $1/r^2$ scaling required by the lensing data).
Thus, the closest-match solutions for $\log_{10} M_b/M_\odot = 8.5$ and $10.0$ are almost identical because they both follow this universal closest match.

However, this is not possible for $\log_{10} M_b/M_\odot = 11.5$ because of the large baryonic mass.
As can be guessed from Fig.~\ref{fig:SFDM-illustrate-lensing}, for such large masses, the $a_{\mathrm{obs}}$ in SFDM either significantly overshoots the observed $a_{\mathrm{obs}}$ at $a_b \sim 10^{-12}\,\mathrm{m/s}^2$ (for relatively large boundary conditions) or the superfluid ends well before the weak-lensing data starts (for relatively small boundary conditions).
In both cases one cannot match to an NFW tail at $a_b \sim 10^{-12}\,\mathrm{m/s}^2$ with an $a_{\mathrm{obs}}$ slightly larger than that of the lensing observations.
For smaller baryonic masses, this can always be achieved by adjusting the boundary condition.

More generally, Fig.~\ref{fig:SFDM-illustrate-best-fit} shows that the closest-match solutions all transition to an NFW tail already at relatively large values of $a_b$.
So the best way to fit the lensing data with SFDM seems to be to just fit an NFW halo to the data.
As mentioned above, NFW halos can -- at best -- explain the weak-lensing RAR when we allow to tune the halo parameters.
For SFDM, this amounts to giving up one of its main motivations.
Namely the idea that SFDM, like MOND, can reproduce the shape and normalization of the RAR without having to tune any boundary conditions or similar parameters.

Of course, in principle, it is possible that galaxy formation provides a mechanism that selects the needed boundary conditions and NFW matching radii with high probability.
However, so far, we have no reason to believe that this is the case, at least not any more or less so than in standard CDM models which also cannot reproduce the observed RAR without galaxy formation picking just the right halo parameters.
Thus, SFDM cannot explain the weak-lensing data without losing its main advantage over standard $\Lambda$CDM models.

Fig.~\ref{fig:SFDM-illustrate-best-fit} also shows  that fitting the weak-lensing data sometimes implies a bad fit of the kinematic data.
An example is the closest match for $\log_{10} M_b/M_\odot = 8.5$ which significantly undershoots the kinematic data.
This is because, for such small masses, reaching the ``universal'' closest match of the lensing data means deviating a lot from the boundary conditions that are optimal for reproducing MOND for kinematic observables.
This can be seen by comparing Fig.~\ref{fig:SFDM-illustrate-best-fit} and Fig.~\ref{fig:SFDM-illustrate-kinematic}.

\section{Discussion and conclusion}

Our analysis used the fiducial numerical parameters proposed by the original authors of SFDM \cite{Berezhiani2018}.
As we argue in appendix~\ref{sec:appendix:parameters}, different choices for these parameters can change the numerical details but cannot avoid the general problem that the model has with gravitational lensing.

Similarly, an improved SFDM model called two-field SFDM was proposed \cite{Mistele2020}.
For this model, the precise numerical details will again be different.
But we expect the same basic problems as in the original SFDM model:
In two-field SFDM, as in the original SFDM model, a MOND-like phonon force is an emergent phenomenon of a superfluid condensate around galaxies.
To avoid constraints from GW170817, this phonon force does not couple to photons.
As a result, the lensing RAR can follow the MOND prediction only when carefully adjusting the boundary condition and NFW matching radius separately for each galaxy.
And even then, unlike in observations, there is no continuity between the kinematic and the lensing RAR.

In conclusion, we have explored whether or not superfluid dark matter can explain the recent data from weak gravitational lensing.
We have found this to be not possible. The major problem is that the observational data continue smoothly from those obtained by
kinematic measurements -- from baryonic matter -- to those obtained from lensing -- from photons. Such a continuation is unexpected in superfluid dark matter because only the baryons couple to the new field, whereas the photons do not.

\acknowledgments

We thank Margot Brouwer, Andrej Dvornik, and Kyle Oman for helpful correspondence.
This work was supported by the DFG (German Research Foundation) under grant number HO 2601/8-1 together with the joint NSF grant PHY-1911909.
This work was supported by the DFG (German Research Foundation) – 514562826.

\appendix

\section{Numerical solutions}
\label{sec:appendix:numerical}

In order to numerically solve Eq.~\eqref{eq:muhateom}, we first split $\hat{\mu}/m$ in a part due only to the baryonic density $\rho_b$, called $\hat{\mu}_b/m$,
\begin{align}
\label{eq:mub}
\Delta\left(-\frac{\hat{\mu}_b}{m}\right) = 4 \pi G \, \rho_b \,,
\end{align}
and the rest, called $\hat{\mu}_{\mathrm{SF}}/m$, due to the superfluid's density,
\begin{align}
\label{eq:muSFeq}
\Delta\left(-\frac{\hat{\mu}_{\mathrm{SF}}}{m}\right) = 4 \pi G \, \rho_{\mathrm{SF}}\left[a_b, \frac{\hat{\mu}_b}{m} + \frac{\hat{\mu}_{\mathrm{SF}}}{m}\right] \,.
\end{align}
As indicated, $\rho_{\mathrm{SF}}$ is a function of $a_b$ and $\hat{\mu}/m$.
This form of $\rho_{\mathrm{SF}}$ can be obtained by plugging the explicit solution of Eq.~\eqref{eq:xeq} into the expression Eq.~\eqref{eq:rhoSF} for $\rho_{\mathrm{SF}}$.
This split of $\hat{\mu}$ into $\hat{\mu}_b$ and $\hat{\mu}_{\mathrm{SF}}$ is not unique since Eq.~\eqref{eq:mub} allows any constant to be added to $\hat{\mu}_b$.
Here, we use
\begin{align}
\frac{\hat{\mu}_b}{m} \equiv \frac{G M_b}{r} \,.
\end{align}

We numerically solve Eq.~\eqref{eq:muhateom} in terms of $u_{\mathrm{SF}} \equiv 10^7 \, \hat{\mu}_{\mathrm{SF}}/m$ using the Julia package `OrdinaryDiffEq.jl` with the `AutoTsit5(Rosenbrock23())` method \citep{Rackauckas2017, Tsitouras2011}.
We use a dimensionless radial variable $y \equiv r/l$ with $l = 1\,\mathrm{kpc}$.
As boundary conditions, we impose
\begin{align}
u_{\mathrm{SF}}'(0) &= 0 \,, \\
u_{\mathrm{SF}}(0) &= \mathrm{const} \,.
\end{align}
The first of these follows from spherical symmetry.
The second corresponds to a choice of chemical potential as discussed above.
To avoid numerical problems at $y \to 0$, we impose these not at $y = 0$ but at $y  = 10^{-5}$, i.e. at $r = 10^{-5}\,\mathrm{kpc}$.

From the numerical solution for $u_{\mathrm{SF}}$ and the analytical solution of $\hat{\mu}_b$, we can directly calculate the accelerations $a_b = G M_b/r^2$ and $a_{\mathrm{SF}}(r) = 10^{-7} u'(y)/l$.
The phonon acceleration $a_\theta$ can be obtained from Eq.~\eqref{eq:atheta} which gives $a_\theta$ in terms of $a_b$ and $x_\beta$.
In particular, we first plug the explicit solution for $x_\beta$ from Eq.~\eqref{eq:xeq} into Eq.~\eqref{eq:atheta}.
This gives $a_\theta$ as a function of $a_b$ and $\hat{\mu}/m = \hat{\mu}_b/m + \hat{\mu}_{\mathrm{SF}}/m$.
Then, we use our numerical solution for $u_{\mathrm{SF}}$ to get the final result for $a_\theta$.

In section~\ref{sec:kinematic}, we introduced the solution for which, given $M_b$, the kinematically-inferred acceleration $a_b + a_\theta + a_{\mathrm{SF}}$ stays within a fraction $\delta$ of MOND up to the largest possible radius $r_{\mathrm{max}}$.
We find this solution, the corresponding boundary condition, and the corresponding $r_{\mathrm{max}}$ by numerically maximizing the radius where a solution first deviates by more than a fraction $\delta$ from MOND using the Julia package `Optim.jl` \citep{Mogensen2018}.

\section{Mass-dependence of deviations from MOND}
\label{sec:appendix:estar}

In Fig.~\ref{fig:SFDM-illustrate-kinematic}, we saw that, for kinematic observables, smaller galaxies follow the MOND prediction more easily than larger galaxies.
To see why, first note that deviations from MOND become relevant when $\varepsilon_*$ violates the condition $\varepsilon_* \ll 1$ (see Eq.~\eqref{eq:mondlimit}).
Consider then the rough analytical lower bound on $\varepsilon_*$ derived in earlier work \cite{Mistele2022},
\begin{align}
 \label{eq:estarlowerbound}
 \varepsilon_* \gtrsim \frac{2 m^2 r}{\alpha M_{\mathrm{Pl}}} + \varepsilon_{*\mathrm{min}} \,,
\end{align}
where $\varepsilon_{*\mathrm{min}}$ is the value of $\varepsilon_*$ that gives $\rho_{\mathrm{SF}} = 0$ (see Eq.~\eqref{eq:rhoSF}, $\varepsilon_{*\mathrm{min}} \approx -0.3$ for the fiducial parameter value $\beta = 2$).
Eq.~\eqref{eq:estarlowerbound} suggests that $\varepsilon_*$ becomes larger than $1$, at the latest, at a universal radius that is independent of baryonic mass $M_b$.
That is, Eq.~\eqref{eq:estarlowerbound} suggests that deviations from MOND set in, at the latest, at this universal radius.
This, in turn, suggests that the maximum radius $r_{\mathrm{max}}$ is roughly independent of $M_b$.
This maximum radius corresponds to a minimum acceleration $a_{b\mathrm{min}} = G M_b/r_{\mathrm{max}}^2$ which is then proportional to $M_b$.
This qualitatively fits with the mass-dependence of deviations from MOND in Fig.~\ref{fig:SFDM-illustrate-kinematic}.

\section{Dependence on parameter values}
\label{sec:appendix:parameters}

The phenomenology of SFDM around galaxies depends on the values of the model parameters $a_0$, $\beta$, and $m^2/\alpha$ \citep{Mistele2022}.
In the main text, we assumed the numerical parameter values proposed by the original authors \cite{Berezhiani2018}.
Here, we discuss the effects of changing these values.

Consider first $a_0$.
In the MOND limit $\varepsilon_* \ll 1$, the kinematically-inferred acceleration $a_{\mathrm{obs}}$ has the form $a_b + \sqrt{a_0 a_b}$.
This roughly reproduces MOND as long as $a_0$ has roughly the value it usually has in MOND, i.e. $a_0 \approx 10^{-10}\,\mathrm{m}/\mathrm{s}^2$.
Thus, we cannot change this parameter much without giving up one of the main ideas behind SFDM, namely to reproduce MOND in this regime.
Thus, for any reasonable value of $a_0$, our conclusions remain the same.

The most important parameter for our purposes is $m^2/\alpha$.
It multiplies both the superfluid energy density $\rho_{\mathrm{SF}}$ as well as the quantity $\varepsilon_*$ that controls whether or not the kinematically-inferred $a_{\mathrm{obs}}$ is in the MOND limit.
With the original fiducial parameter values \cite{Berezhiani2018} and at large $a_b$,
the SFDM prediction for the
lensing-inferred $a_{\mathrm{obs}}$ is much smaller than observations
(see section~\ref{sec:weaklensing}).
To counter this, we could choose a larger value of $m^2/\alpha$ which gives a larger superfluid mass and thus a larger lensing-inferred $a_{\mathrm{obs}}$.
However, this also makes $\varepsilon_*$ larger.
As a result, the kinematically-observed $a_{\mathrm{obs}}$ starts to deviate from the MOND prediction already at larger $a_b$.

\begin{figure}
\centering
\includegraphics[width=.8\textwidth]{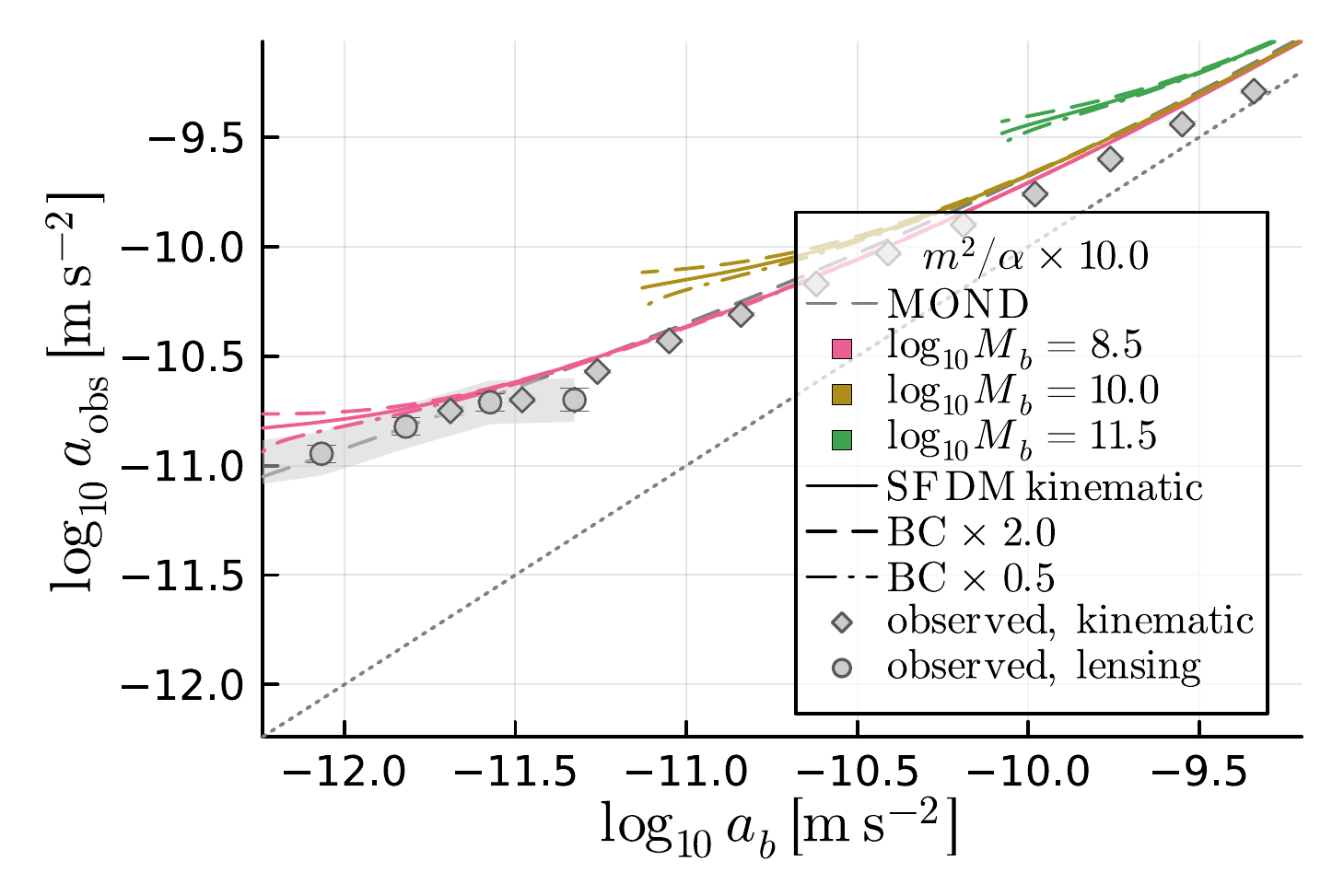}
\caption{
    Same as Fig.~\ref{fig:SFDM-illustrate-kinematic} but with $m^2/\alpha$ increased by a factor $10$.
}
\label{fig:SFDM-illustrate-kinematic-fm2-10}
\end{figure}

\begin{figure}
\centering
\includegraphics[width=.8\textwidth]{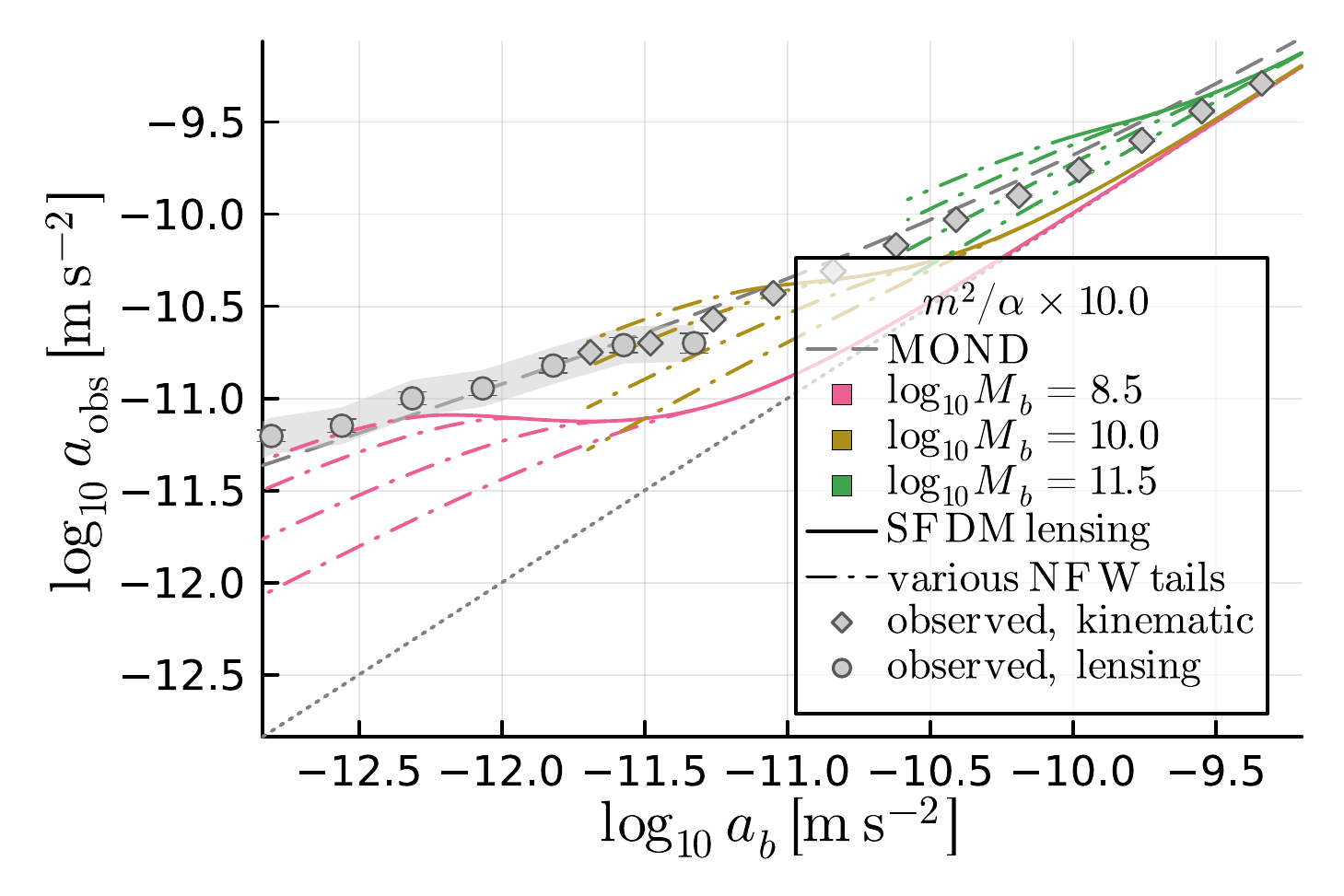}
\caption{
    Same as Fig.~\ref{fig:SFDM-illustrate-kinematic-vs-lensing-NFWtail} but with $m^2/\alpha$ increased by a factor $10$.
}
\label{fig:SFDM-illustrate-kinematic-vs-lensing-NFWtail-fm2-10}
\end{figure}

\begin{figure}
\centering
\includegraphics[width=.8\textwidth]{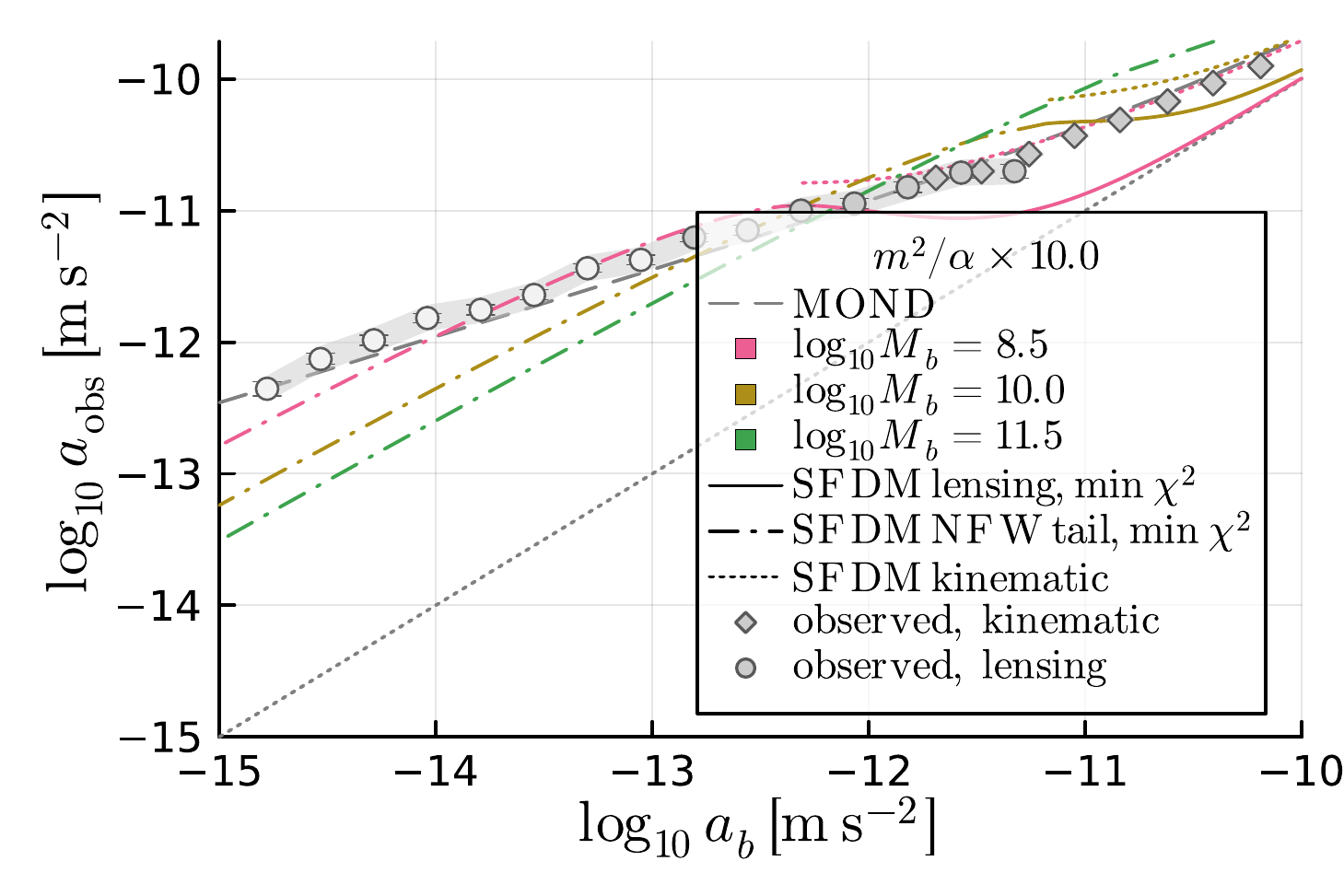}
\caption{
    Same as Fig.~\ref{fig:SFDM-illustrate-best-fit} but with $m^2/\alpha$ increased by a factor $10$.
    The reduced $\chi^2$ values for the closest-match solutions shown here are $13.6$, $43.8$, and $66.9$ for $\log_{10} M_b/M_\odot = 8.5$, $10.0$, and $11.5$, respectively.
    Note that the ``universal'' closest match described in Sec.~\ref{sec:weaklensing:quantitative} is the same as before, but now it can be reached only for the lowest baryonic mass shown here, $\log_{10} M_b/M_\odot = 8.5$.
    Due to the increased $m^2/\alpha$, the other masses are too large to reach it.
    This is why all three closest-match solutions shown here are different while in Fig.~\ref{fig:SFDM-illustrate-best-fit} two of them are almost identical.
}
\label{fig:SFDM-illustrate-best-fit-fm2-10}
\end{figure}

That is, by adjusting the parameter $m^2/\alpha$ we can change where the problems discussed in section~\ref{sec:weaklensing} occur.
But we cannot avoid these fundamental problems of SFDM.
This is illustrated in Fig.~\ref{fig:SFDM-illustrate-kinematic-fm2-10}, Fig.~\ref{fig:SFDM-illustrate-kinematic-vs-lensing-NFWtail-fm2-10}, and Fig.~\ref{fig:SFDM-illustrate-best-fit-fm2-10}.
These are the same as Fig.~\ref{fig:SFDM-illustrate-kinematic}, Fig.~\ref{fig:SFDM-illustrate-kinematic-vs-lensing-NFWtail}, and Fig.~\ref{fig:SFDM-illustrate-best-fit} but with $m^2/\alpha$ increased by a factor $10$.
We see that there is still a tension between the kinematic and the lensing RAR.
And reproducing a MOND-like lensing RAR is, at best, possible when tuning the boundary condition and NFW matching radius.

\begin{figure}
\centering
\includegraphics[width=.8\textwidth]{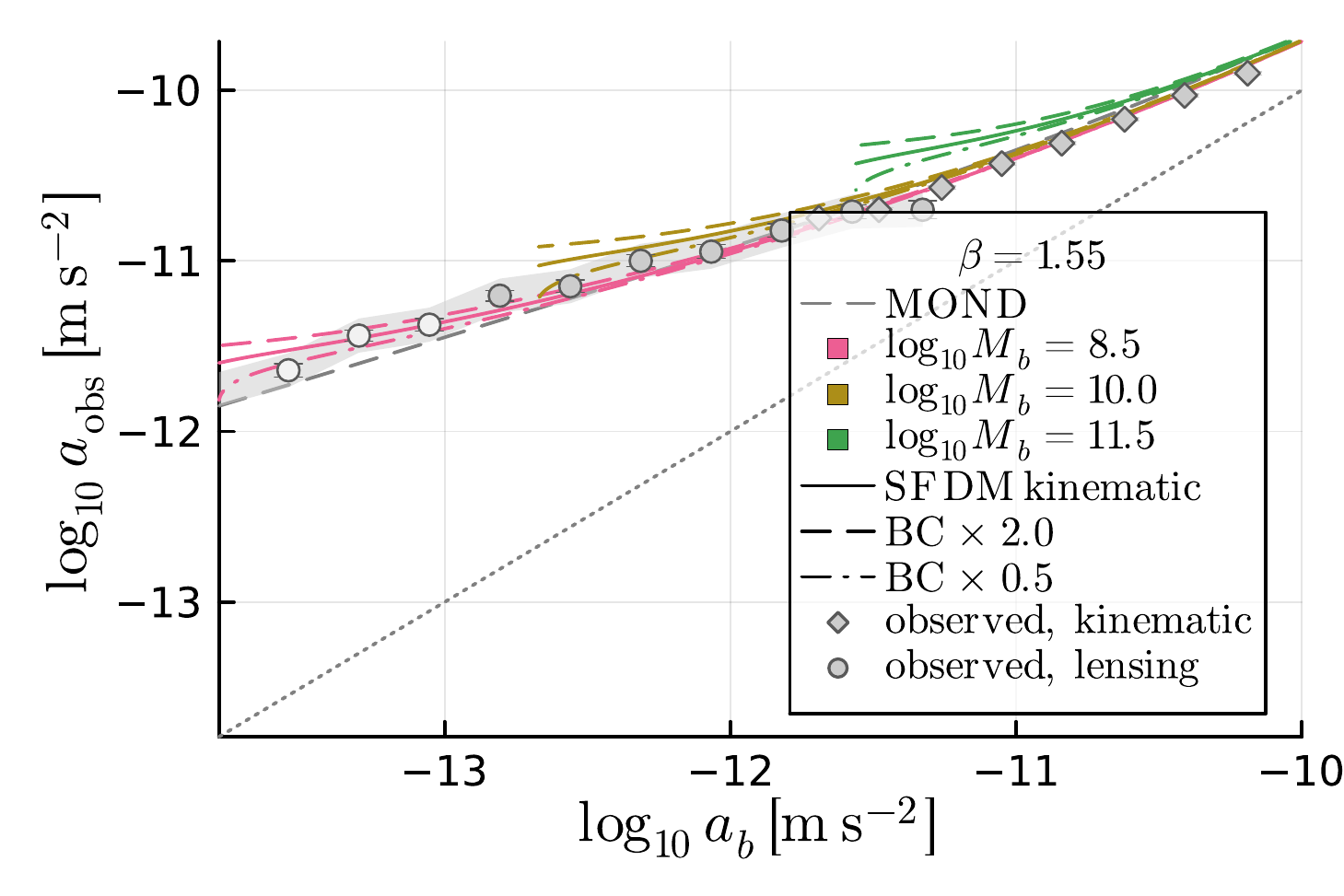}
\caption{
    Same as Fig.~\ref{fig:SFDM-illustrate-kinematic} but with $\beta = 1.55$ instead of $\beta = 2$.
}
\label{fig:SFDM-illustrate-kinematic-beta-1.55}
\end{figure}

\begin{figure}
\centering
\includegraphics[width=.8\textwidth]{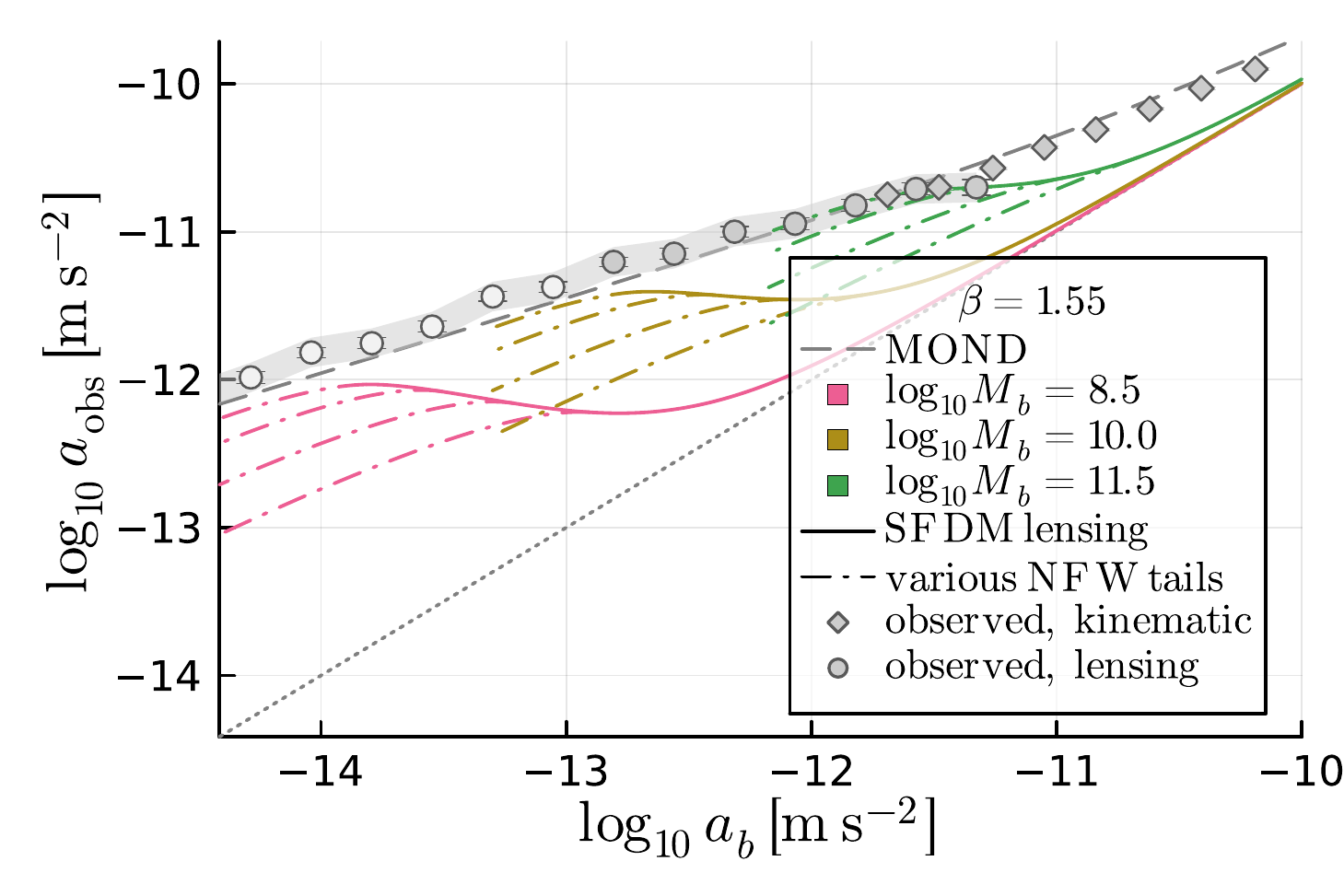}
\caption{
    Same as Fig.~\ref{fig:SFDM-illustrate-kinematic-vs-lensing-NFWtail} but with $\beta = 1.55$ instead of $\beta = 2$.
}
\label{fig:SFDM-illustrate-kinematic-vs-lensing-NFWtail-beta-1.55}
\end{figure}

\begin{figure}
\centering
\includegraphics[width=.8\textwidth]{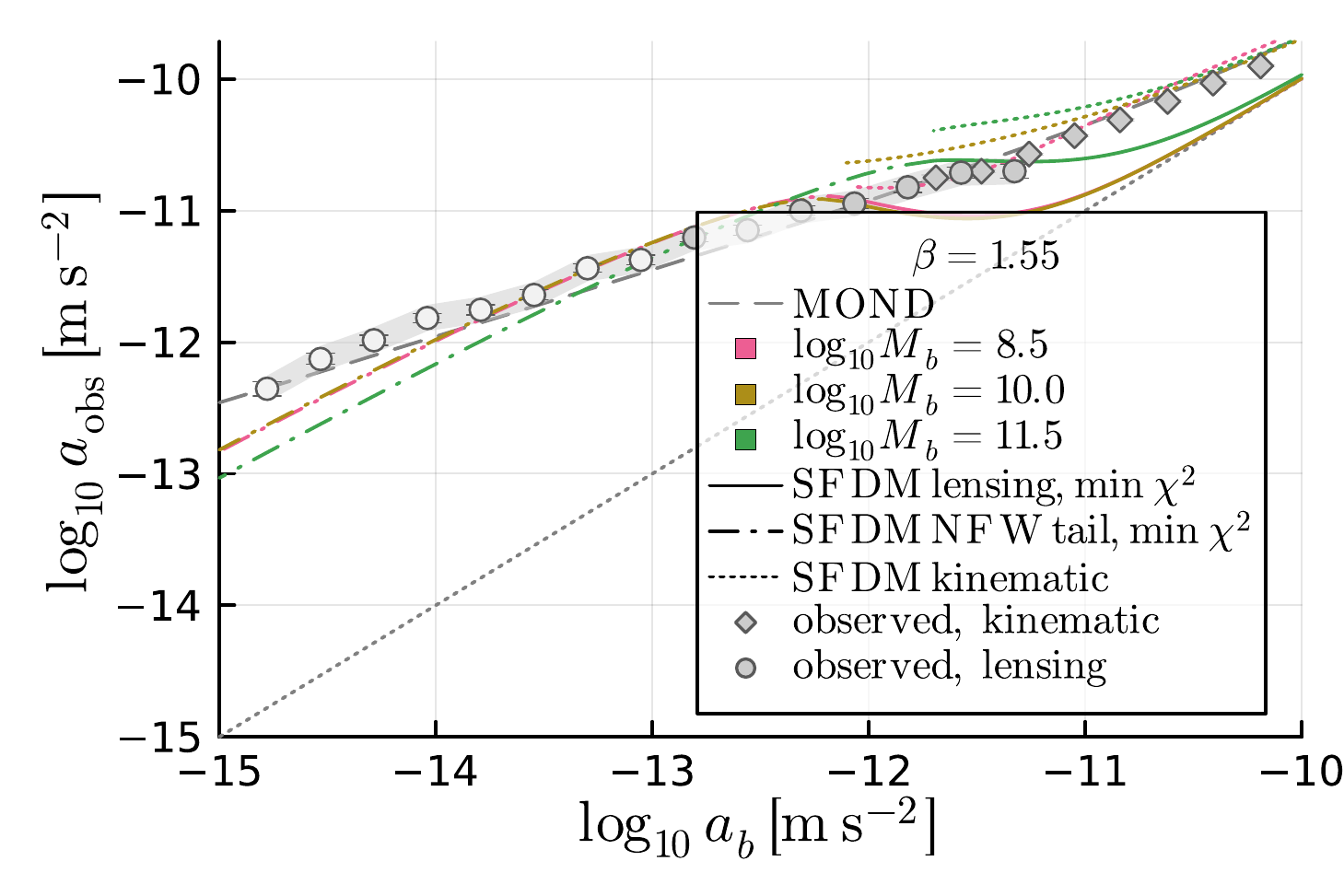}
\caption{
    Same as Fig.~\ref{fig:SFDM-illustrate-best-fit} but with $\beta = 1.55$ instead of $\beta = 2$.
    The reduced $\chi^2$ values for the closest-match solutions shown here are $15.0$, $14.9$, and $24.5$ for $\log_{10} M_b/M_\odot = 8.5$, $10.0$, and $11.5$, respectively.
}
\label{fig:SFDM-illustrate-best-fit-beta-1.55}
\end{figure}

This leaves $\beta$.
This parameter controls the precise shape of $\rho_{\mathrm{SF}}$ and the phonon force $a_\theta$ outside the MOND limit, i.e. for $\varepsilon_* \gtrsim 1$ (see for example Appendix A of \cite{Mistele2022}).
To avoid instabilities and to ensure a positive superfluid density, $\beta$ must be between $3/2$ and $3$ \citep{Berezhiani2015}.
In Fig.~\ref{fig:SFDM-illustrate-kinematic-beta-1.55}, Fig.~\ref{fig:SFDM-illustrate-kinematic-vs-lensing-NFWtail-beta-1.55}, and Fig.~\ref{fig:SFDM-illustrate-best-fit-beta-1.55}, we show the effect of using $\beta = 1.55$ instead of the original fiducial value $\beta = 2.0$ \cite{Berezhiani2015}.
We see that there are minor quantitative changes but our qualitative conclusions remain unchanged.

Even if changing $\beta$ did lead to significant changes, it would be best to not rely on them too much.
This is because both the value of $\beta$ and the form of the finite-temperature corrections that $\beta$ is supposed to represent are completely ad-hoc \citep{Mistele2020}.
So any significant changes related to $\beta$ might easily turn out to be unphysical.

\section{Upper bound on lensing $a_{\mathrm{obs}}$ as a function of $r_{\mathrm{NFW}}$}
\label{sec:appendix:aDMupperbound}

Suppose we are given a solution of the equations of motion inside the superfluid core of a galaxy.
This does not fix the acceleration outside the superfluid since this depends on where exactly we match the superfluid density to an NFW $1/r^3$ tail.

There is an optimal $r_{\mathrm{NFW}}$ that maximizes the acceleration outside the superfluid, as we will now show.
Using the condition that the superfluid density $\rho_{\mathrm{SF}}$ matches the NFW tail's density at $r = r_{\mathrm{NFW}}$, we have outside the superfluid core for the acceleration $a_{\mathrm{DM}}$ due to dark matter,
\begin{align}
a_{\mathrm{DM}}(r) = \frac{G (M_{\mathrm{SF}}(r_{\mathrm{NFW}}) + 4 \pi \rho_{\mathrm{SF}}(r_{\mathrm{NFW}}) r_{\mathrm{NFW}}^3 \ln(r/r_{\mathrm{NFW}}))}{r^2} \,.
\end{align}
We find the maximum by differentiating with respect to $r_{\mathrm{NFW}}$ and setting the result to zero, which gives
\begin{align}
\label{eq:sfdm:appendix:aDMmax}
0 = \rho_{\mathrm{SF}}'(r_{\mathrm{NFW}}) r_{\mathrm{NFW}} + 3 \rho_{\mathrm{SF}}(r_{\mathrm{NFW}})\,,
\end{align}
where we used $M_{\mathrm{SF}}'(r) = 4 \pi r^2 \rho_{\mathrm{SF}}(r)$.
Note that this is independent of $r$.
Thus, there is a single $r_{\mathrm{NFW}}$ that maximizes the acceleration everywhere outside the superfluid.

In earlier work \cite{Mistele2022}, we showed that -- for a given solution inside the superfluid -- the virial mass $M_{200}$ has a maximum as a function of $r_{\mathrm{NFW}}$.
It turns out the $r_{\mathrm{NFW}}$ that maximizes $M_{200}$ discussed there is the same $r_{\mathrm{NFW}}$ that maximizes $a_{\mathrm{DM}}$ discussed above.
To see this, consider the virial mass $M_{200} = (4\pi/3) \rho_{200} r_{200}^3$ where $\rho_{200} = 200 \cdot 3 H^2/(8 \pi G)$ with the Hubble constant $H$.
From its definition it's clear that $M_{200}$ is maximal when $r_{200}$ is maximal.
For $r_{200}$, we have
\begin{align}
\frac{M_{\mathrm{SF}}(r_{\mathrm{NFW}}) + 4 \pi \rho_{\mathrm{SF}}(r_{\mathrm{NFW}}) r_{\mathrm{NFW}}^3 \ln(r_{200}/r_{\mathrm{NFW}})}{\frac{4\pi}{3} r_{200}^3} = \rho_{200} \,,
\end{align}
where we consider $r_{200}$ as a function of $r_{\mathrm{NFW}}$, i.e. $r_{200} = r_{200}(r_{\mathrm{NFW}})$
Let's differentiate both sides with respect to $r_{\mathrm{NFW}}$.
This gives
\begin{align}
0 = \frac{\partial a_{\mathrm{DM}}(r)}{\partial r_{\mathrm{NFW}}}\rvert_{r = r_{200}(r_{\mathrm{NFW}})} + \frac{\partial r_{200}(r_{\mathrm{NFW}})}{\partial r_{\mathrm{NFW}}} \cdot ( \dots )\,,
\end{align}
where we leave out the terms multiplying $\partial_{r_{\mathrm{NFW}}} r_{200}$ for brevity.
Indeed, at the maximum $r_{200}$, this derivative vanishes, $\partial_{r_{\mathrm{NFW}}} r_{200} = 0$.
The remaining term is equivalent to Eq.~\eqref{eq:sfdm:appendix:aDMmax}.
Thus, the maximum of $r_{200}$ and the maximum of $a_{\mathrm{DM}}(r)$ outside the superfluid are reached for the same $r_{\mathrm{NFW}}$.

In extreme cases, it can happen that the virial radius $r_{200}$ is inside the superfluid core.
In this case, $r_{200}$ and $a_{\mathrm{DM}}$ are not maximized by the same $a_{\mathrm{DM}}$.
For definiteness, above we numerically maximize $a_{\mathrm{DM}}$ and not $r_{200}$.
Concretely, we use the Julia package `Optim.jl` \citep{Mogensen2018} to find which $r_{\mathrm{NFW}}$ maximizes $a_{\mathrm{DM}}$ for $r_{\mathrm{NFW}}$ in the interval $[0.001\,\mathrm{kpc}, r_0]$ where $r_0$ is the radius where the superfluid density drops to zero.

\bibliographystyle{JHEP}
\bibliography{SFDM-lensing.bib}

\end{document}